\documentclass[12pt,preprint]{aastex}

\slugcomment{}

\shorttitle{Decelerating Flows in  TeV Blazars}
\shortauthors{Georganopoulos \& Kazanas}

\begin{document}

\title{Decelerating Flows in  TeV Blazars: A Resolution to the\\
BL Lac -- FR I Unification Problem}

\author{Markos Georganopoulos\altaffilmark{1}
 \& Demosthenes Kazanas\altaffilmark{2}}
\affil{Laboratory for High Energy Astrophysics, NASA Goddard Space Flight Center, 
Code 661, Greenbelt, MD 20771, USA.}
\altaffiltext{1}{Also NAS/NRC Research Associate; email:
markos@milkyway.gsfc.nasa.gov}
\altaffiltext{2}{email: Demos.Kazanas-1@nasa.gov}

%\authoraddr{}
%\email{}

\begin{abstract}

TeV emission from BL Lacertae (BL) objects is commonly modeled as 
Synchrotron-Self Compton (SSC) radiation from relativistically 
moving homogeneous plasma blobs. In the context of these models, 
the blob Lorentz factors needed to reproduce the corrected for 
absorption by the diffuse IR background (DIRB) TeV emission are large ($\delta 
\gtrsim  50$). The main reason for this is that stronger beaming  
eases the problem of the lack of  $\sim$ IR-UV synchrotron seed 
photons needed to produce the de-absorbed $\sim $ few TeV peak 
of the spectral energy distribution (SED). However, such  high 
Doppler  factors are  in strong disagreement with the unified scheme, 
according to which  BLs are FR I radio galaxies with their jets 
closely aligned to the line of sight. Here, motivated by the 
detection of sub-luminal velocities in the sub-pc scale jets of 
the best studied TeV blazars, MKN 421 and MKN 501, we examine the 
possibility  that the relativistic flows in the TeV BLs decelerate. 
In this case, the problem of the missing seed photons is solved 
because of Upstream Compton (UC) scattering,  a process in which  
the upstream energetic electrons from the fast base of the flow `see' 
the synchrotron seed photons produced in the slow part of the flow 
relativistically beamed. Modest Lorentz factors ($\Gamma \sim 15$), 
decelerating down to  values compatible with the recent radio 
interferometric observations,  reproduce the  $\sim $ few TeV peak energy
of these sources. Furthermore, such  decelerating flows are shown 
to be in  agreement with the BL - FR I unification, naturally reproducing
the observed BL/FR I broad band luminosity ratios. 
\end{abstract}


\section{Introduction}

There is a small but growing family of blazars detected at TeV energies. 
These belong exclusively to the class of high peak frequency BLs, 
i.e. blazars whose synchrotron component peaks at X-ray energies.
TeV emitting BLs are  of particular interest because of the possibility of absorption of their TeV emission by the DIRB
\citep{nikisov62, gould66, stecker92}. Study 
of their spectra in the TeV range can  be used to probe the properties of 
DIRB as a function of redshift $z$ \citep{salamon98}, 
given that the magnitude of absorption 
depends on the redshift of the source and the, still elusive, DIRB spectrum  \citep{malkan01,primack01,aharonian02}.

The absorption of the TeV photons of this blazar class  suggests that
 both  the  
intrinsic peak photon energy $E_p$ and peak luminosity $L_p$ of the high 
energy (TeV) component are higher than those observed. Even for the nearby 
($z=0.031$) MKN 421, $E_p$ can increase by a factor of $\sim 10$ after 
de-absorption to $\sim 5-10 $ TeV \citep{dejager02}. The de-absorbed spectrum 
of   H1426+428 at z=0.129 is  even more extreme, characterized  by $E_p 
\gtrsim 10$  TeV \citep{aharonian02}. 
Modeling of these sources has been done in the framework of the homogeneous
SSC model  [e.g.  \cite{coppi92, mastichiadis97}], 
according to which a blob of energetic plasma is moving with a constant
Lorentz factor $\Gamma$ forming  a small angle $\theta$ to the line of sight.
Such models require high Doppler factors ($\delta =1/\Gamma(1-
\beta\cos\theta)\gtrsim 50$,  where $\beta$ is the dimensionless speed 
of the flow and $\theta$ its angle to the observer's line of sight) to 
reproduce the de-absorbed $E_p$  [e.g. \cite{krawczynski02}; 
see also next section]. However, even smaller values of 
$\delta (\simeq 10)$ are in conflict \citep{chiaberge00} with the unification 
scheme according to which BLs represent FR I radio galaxies viewed at small
$\theta \, (\simeq 1/\Gamma$) \citep{urry95}. Also, these high values of
$\delta$ are in disagreement with the small values of the apparent 
velocities observed in the sub-pc regions of the TeV BL Mkn 421 and Mkn 501
(e.g. \cite{marscher99}).
In this note we propose that the above issues can be resolved by postulating
that the TeV blazar emission originates in a relativistic but {\sl decelerating}
flow. In \S 2 we present a quantitative analysis and formulation of the above 
arguments, while in \S 3 we outline the basic notions behind our proposal 
and explain why and how they resolve the outstanding issues discussed in 
\S 2. Finally, in \S 4 we discuss some further issues.

\section{Problems with Uniform Velocity TeV Blazar Models}

{\bf The Blazar Spectra:} One of the characteristics of the synchrotron 
components of the TeV blazar spectra is a break at an energy $\epsilon_b 
\sim 10^{-4} - 10^{-6}$ (unprimed energies are in the observer frame while 
primed ones in the flow rest frame, all normalized to the rest mass of 
the electron $m_ec^2$), with most of the (comoving) 
synchrotron energy density above $\epsilon'_b$, a feature that significantly
affects their  TeV emission:  Because of the reduction in the inverse Compton
(IC) scattering cross section in the K-N regime and the break in the photon
energy density at $\epsilon' < \epsilon'_b$, electrons with energies
$\gamma \gtrsim 1/ \epsilon'_b$ will channel a decreasing fraction of 
their energy to IC scattering, leading to a peak in the IC luminosity at 
$\epsilon'_p \simeq 1/\epsilon'_b$ %(so that $\epsilon'_p \, \epsilon'_b \simeq 1$) 
even if the maximum electron energy is $\gamma_{max} \gg 1/\epsilon'_b$. 
For a source moving with a Doppler factor $\delta$ relative to the 
observer $\epsilon'_{b}$ and $\epsilon'_p$ will be $\epsilon_{b}=\delta
\epsilon'_{b}$ and $\epsilon_{p}=\delta\epsilon'_{p}$ yielding 
\begin{equation}
\delta^2 (\epsilon'_b \, \epsilon'_{p}) \simeq (\epsilon_b \, \epsilon_p)
~~{\rm or}~~ \delta \simeq (\epsilon_{b}\, \epsilon_{p})^{1/2} =   40\;(\nu_{b,16} \;E_{p,\,10\,\rm TeV})^{1/2},  \label{d_constr}
\end{equation} 
where $\nu_{b,16}$ is the {\sl observed} synchrotron break frequency in units
 of $10^{16}$ Hz and $E_{p,\,10\,\rm TeV}$ is the energy of the {\sl 
de-absorbed} IC peak  in units of 10 TeV. De-absorbed $E_p$ values in excess 
of 10 TeV then imply relativistic flows in blazars with $\Gamma \gtrsim 40$.
The crucial point in the above argument, namely that the IC luminosity  
peaks at $\epsilon'_p\lesssim 1/ \epsilon'_b$, can be demonstrated explicitly
within the homogeneous SSC models: Assume, as customary, continuous injection 
of a power law electron distribution within a uniform source at a rate $Q(\gamma)
\propto \gamma^{-s}$, $\gamma \leq \gamma_{max}$.  The steady state electron 
distribution is then
\begin{equation}
n(\gamma)\propto\left\{\begin{array}{ll}
                  \gamma^{-s} & \mbox{for $\gamma < \gamma_b$}\\
		  \gamma^{-(s+1)} & \mbox{for $ \gamma_b\leq \gamma \leq \gamma_{max}$,}
		  \end{array}
	\right.		\label{n(g)}
\end{equation}
with $\gamma_b$ the electron energy below which electrons escape from the
source faster than they radiatively cool. The corresponding comoving 
synchrotron  energy density distribution is
\begin{equation}
u(\epsilon')\propto\left\{\begin{array}{ll}
                  \epsilon'^{-(s-1)/2} & \mbox{for $\epsilon' < \epsilon'_b$}\\
		  \epsilon'^{-s/2} & \mbox{for $ \epsilon'_b\leq \epsilon \leq \epsilon'_{max}$,}
		  \end{array}
	\right.	 \label{u(e)}	
\end{equation}
where $\epsilon'_b=b \gamma_b^2$,  $\epsilon'_{max}=b \gamma_{max}^2$,
and $b$ is the comoving magnetic field in units of its critical value 
$B_c=m_e^2c^3/e\hbar=4.4 \times 10^{13}$ G. Fits to the synchrotron
spectra of TeV blazars require $1 < s < 2$, with comoving peak synchrotron 
luminosity at $\epsilon'_{max}$. We now examine the energy $\epsilon'_p$ 
at which the IC luminosity peaks as a function of the maximum electron 
energy $\gamma_{max}$. The K-N influence on the cross section begins at 
$\gamma_{max} \simeq 1/\epsilon'_{max}$. Above that energy the electrons 
interact only with the fraction of the synchrotron spectrum at energies 
less than $\epsilon' \lesssim 1/\gamma$, while the maximum photon energy
resulting from the IC is $\epsilon'_M \approx \gamma_{max}$. If 
$L(\epsilon'_M)$ is the photon scattering rate to energy $\epsilon'_M$, 
the IC luminosity at this energy is
\begin{equation}
\epsilon'_M L(\epsilon'_M)\propto \epsilon'_M \; n(\gamma_{max})
\;  \epsilon'    u(\epsilon')   \; \gamma_{max}^2.  
\end{equation} 
Setting $\epsilon'=1/\gamma_{max}$ as the appropriate seed photons (photons
of larger energy are in the K-N regime, and photons of lower energy give lower
 $\epsilon'_{IC}$), and using eq. (\ref{n(g)}), (\ref{u(e)}) we obtain
\begin{equation}
\epsilon'_M L(\epsilon'_M)\propto
\left\{\begin{array}{ll}
                  \epsilon'^{(2-s)/2}_M & \mbox{for $\epsilon'_M \lesssim 1/\epsilon'_b$}\\
\\
		  \epsilon'^{(1-s)/2}_M & \mbox{for $\epsilon'_M \gtrsim 1/\epsilon'_b$}
		  \end{array}
	\right.
\end{equation} 
where we have also used  $\epsilon'_M=\gamma_{max}$. Therefore,
for $1<s<2$  the luminosity at 
maximum photon energy $\epsilon'_M L(\epsilon'_M)$ increases with 
$\gamma_{max}$ for $\epsilon'_M=\gamma_{max}\lesssim 1/\epsilon'_b$ and
decreases for $\epsilon'_M \gtrsim 1/\epsilon'_b$, achieving its peak
luminosity at energy $\epsilon'_p\approx 1/\epsilon'_b$.

{\bf Blazar Unification:} According to the unification scheme of radio 
loud active galaxies (e.g. Urry  \& Padovani 1995) BLs are  FR I 
radio  galaxies with their jets oriented close to the line of sight.
The average Lorentz factor $\Gamma$ of 
the jet flows, derived by matching the luminosity functions of  
BL and FR I samples, were estimated to be $\Gamma\sim3-5$ \citep{urry91,
hardcastle03}, in clear disagreement with the values of the Doppler factors 
required by the homogeneous SSC models for the TeV blazars. The high
Doppler factors estimated on the basis of homogeneous SSC models imply
that for $\Gamma \simeq \delta \simeq 50$, $\theta \approx 1/\Gamma\approx
1^{\circ}$ requiring sources very well aligned to the line of sight, thus
grossly overpredicting the number of FR I galaxies above a given limiting
flux (this actually would be the case even with the much smaller value of 
$\Gamma \simeq 10$; \cite{hardcastle03}). 

In a different aspect of the same problem, Chiaberge et al. (2000) showed 
that the FR I nuclei are overluminous by a factor of $10-10^4$ compared 
to their luminosity should they have been misaligned BLs harboring flows with 
Lorentz factors $\Gamma \sim 15$. Applying to sub-pc scales the arguments
of Laing et al. (1999) concerning the structure of FR I kpc scale jets, 
they opted for jets with a high $\Gamma$ `spine' surrounded by lower 
$\Gamma$ sheath. For a source at a small angle to the line of sight the 
emission is dominated by the fast spine, while, at large angles this  
radiation is beamed out of the observer's direction and the observed
spectrum is dominated by the mildly beamed emission by the slower sheath.

However, recent VLBA \citep{marscher99}, VLBI \citep{edwards02}, 
and combined VSOP and VLBI \citep{piner99} studies do not detect any 
high velocity components in the jets of the two TeV sources MKN 421 
and MKN 501. These observations are compatible  with  subluminal
($\beta_{app} \sim0.3-0.6$, Piner et al. 1999, Edwards et al. 2002) or
mildly relativistic ($\beta_{app}\sim 2$, Marscher 1999) sub-pc 
velocities.  A value of $\delta\sim 50$, as needed  in modeling the TeV
emission of these sources, could produce the observed velocities only
for $\theta \lesssim 0.1^{\circ}$. Rather than assuming such an 
extraordinary jet alignment for both sources, Marscher (1999) suggested 
that the flow in the sub-pc environment of these sources has already 
decelerated substantially.  Additional support for slow  flows in  sub-pc
scales comes from   \cite{jorstad01} that showed that in 
several cases VLBI components in BLs move with $\beta_{app}\sim 1-2$.
That low jet velocities at pc scales are real and not the result of 
projection effects is supported by the observation of subluminal velocities
at the jet of the FR I galaxy 3C 270 jets, which are thought to be at 
large angle to the observer's line of sight (Piner et al. 2001).

\section{Decelerating flows and UC emission}

Motivated by the above issues,  we propose that 
in the  high energy emitting  region  of the  TeV BLs the plasma flow is 
 relativistic and {\sl decelerating} (a similar proposal was advanced to unify
the broadband properties of the hot spots of FR II radio 
galaxies and quasars \citep{georganopoulos03}).
Our proposed scheme for the BL flows involves the injection of a power law 
electron distribution at the base of a relativistic flow which decelerates
while at the same time the electron distribution cools radiatively. 
The highest synchrotron frequencies  originate 
 at fast base of the flow where the 
electrons are more energetic. As both 
the flow velocity and electron  energy drop with radius, the locally emitted 
synchrotron spectrum shifts to lower energies while its beaming pattern 
becomes wider. At small angles the observed spectrum is  dominated by emission 
from the higher $\Gamma$ base of flow, where the most energetic electrons
reside. At larger angles this emission from the inner, fast flow section 
is beamed away from the observer and the major contribution to the 
spectrum comes from its slower parts which contain less energetic
electrons, leading to softer spectra.   

The inverse Compton emission of such a flow behaves in a more involved
way: Electrons will upscatter the locally produced synchrotron seed photons,
giving rise to a local SSC emission with $\delta-$dependence similar to 
   that of synchrotron. However, the electrons of a given radius scatter 
will  also those synchrotron photons produced 
downstream in the flow. The energy density of the latter, will appear 
Doppler boosted in the fast (upstream) part of the flow by $\sim 
\Gamma_{rel}^2$ \citep{dermer95}, where $\Gamma_{rel}$ is the relative 
Lorentz factor between the fast and slow part of the flow. With their
maximum energy being lower (because of cooling) and their energy density
amplified they  contribute to the IC emission at energies higher 
than expected on the basis of uniform velocity models without the need of
invoking as large Doppler factors.
The beaming pattern of this UC radiation is also intermediate
between the  synchrotron/SSC pattern of $\delta^{2+\alpha}$ and the external
Compton pattern (EC) of  $\delta^{3+2\alpha}$  \citep{dermer95, georganopoulos01}, where $\alpha$ is
 the spectral index of the radiation. To demonstrate this consider
a two-zone flow, a fast part with Lorentz factor $\Gamma_1$ followed by a
 slower
part with Lorentz factor $\Gamma_2$. Consider also an observer
located at an angle $\theta$ such that  the  Doppler factors of the
two zones are $\delta_1 $, $ \delta_2$.  The beaming pattern of the UC 
radiation in the frame of the slow part of the flow will be 
$\delta_{1,2}^{3+2\alpha}$, where 
$\delta_{1,2}$ is the Doppler factor of the fast flow in the frame of the
slow flow. To convert this beaming pattern to the observer's frame
we need to boost it by $\delta_2^{2+\alpha}$. The beaming pattern is then
written as $\delta_{1,2}^{3+2\alpha}\,\delta_2^{2+\alpha}$. To write
$\delta_{1,2}$ as a function of $\delta_{1}$, $\delta_{2}$, we note that
a photon emitted in the fast part of the flow is seen by the observed
boosted in energy by a factor $\delta_1$. The same boosting can take place
is two stages: first going to the frame of the slow flow by being boosted 
by  $\delta_{1,2}$ and then going to the observer's frame by  being boosted 
by $\delta_2$. Because  the final photon energy in the observer's frame does
not depend on the intermediate transformations,  $\delta_{1,2}=\delta_{1}
/\delta_{2}$.  The beaming pattern of UC scattering is therefore 
$\delta_1^{3+2\alpha}/\delta_2^{1+\alpha}$.   Note that, as expected, 
 for $\delta_1=\delta_2$, we recover the beaming pattern of SSC, 
while for $\delta_2=1$, that of EC radiation.

%\subsection{Applications to TeV Blazars}

\begin{figure}
\epsscale{0.7}
\plotone{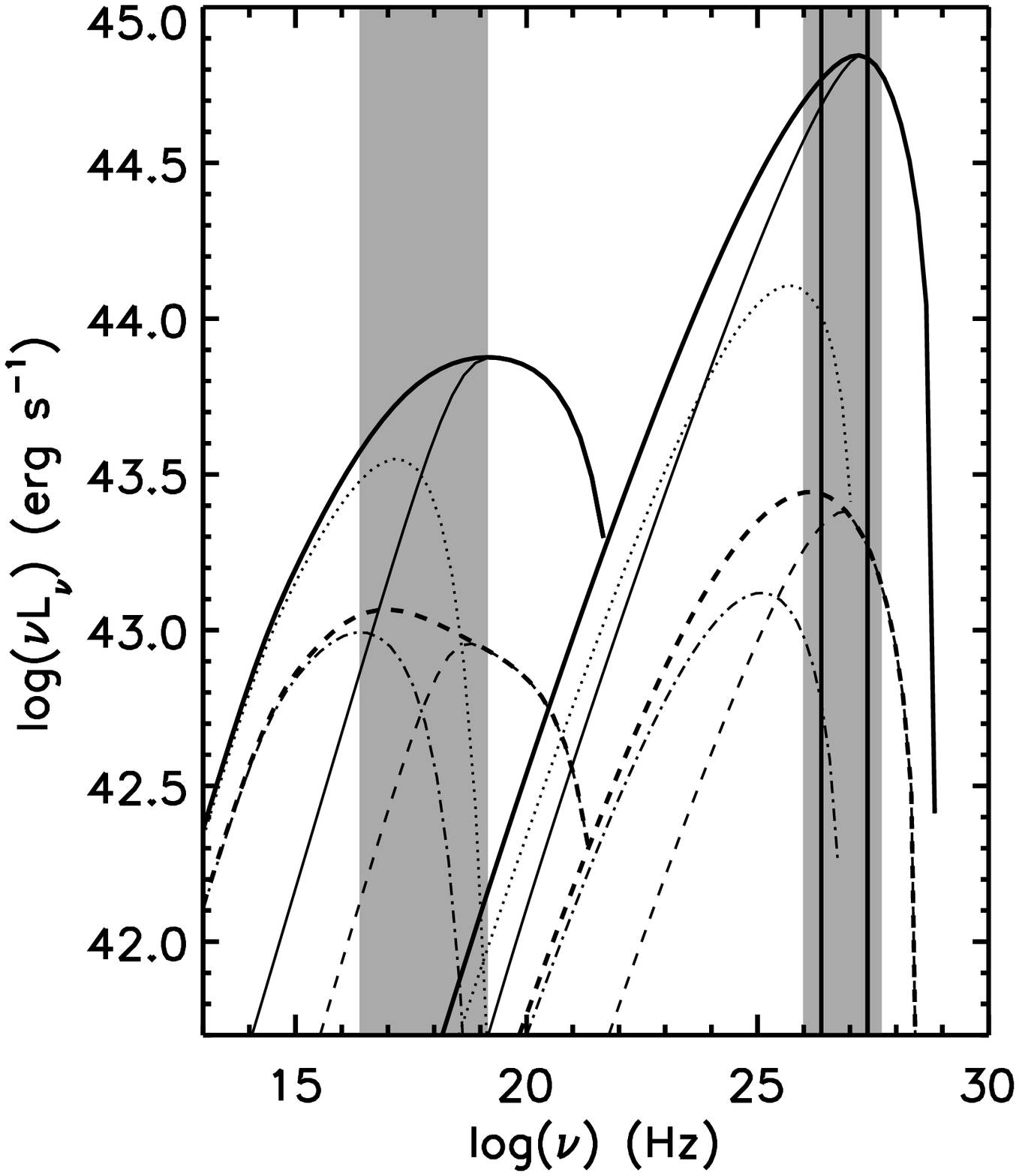}
\caption{The synchrotron and inverse Compton emission from a decelerating
 relativistic flow under $\theta=3^o$ (thick solid line) 
and $\theta=6^o$ (thick broken line)
 observing angles. The flow decelerates from $\Gamma_1=15$ to $\Gamma_2=4$
 within a length  $Z=2\times 10^{16}$ cm. The radius of the cylindrical
 flow is set to $R=Z=2\times 10^{16}$ cm. 
A power law electron energy distribution,  $n(\gamma)\propto \gamma^{-2}$,
$\gamma \leq 3\times 10^7$ is injected at the base of the flow with
a  magnetic field $B=0.1$ G,  half  of the   equipartition value.
The thin solid and broken lines correspond to the emission due to the
fast inner $10 \%$  of the flow, while the dotted and dash-dotted thin lines
correspond to the rest of the flow. 
The shaded areas correspond approximately to the energy 
range of X-ray and TeV telescopes and the two vertical solid lines energies
to  1 and 10 TeV.}
\label{fig1}
\end{figure}

To demonstrate the relevance of  decelerating flows in TeV blazars, we 
developed a simple model, based on an one dimensional kinematic flow 
description. An electron distribution $n(\gamma)\propto \gamma^{-2}$ is 
injected at the base of a decelerating relativistic flow  with velocity 
profile $\Gamma(z)=\Gamma_o (z/z_o)^{-2}$. Electrons cool radiatively
as they propagate downstream. We calculate their radiative losses along 
the flow and also their energy distribution as a function of $z$. We then 
calculate the synchrotron emissivity along $z$ and, performing the necessary 
beaming transformations and $z-$integration, the volume integrated synchrotron 
emission  as a function of observing angle $\theta$. Using the synchrotron 
emissivity  as a function of $z$ we calculate the SSC and UC emissivities
as a function of $z, \, \theta$. A final integral over $z$ then provides
the volume integrated  Compton emissivity as a function of $\theta$.
In fig. \ref{fig1} we plot the SED for a deceleratring flow for two different
observing angles. Note that at  $\theta=3^{\circ}$ this model 
achieves a peak energy for the high energy  component at $\sim 10$ TeV,
 using a modest Lorentz 
factor of $\Gamma_1=15$. Note also  the stronger angle dependence of
emission from the inner fast part of the flow which produces the
highest frequencies in each spectral component.
Finally, note that,
 in contrast 
to single velocity homogeneous SSC models,  the
Compton component is more sensitive to orientation than synchrotron,
as expected if  UC scattering dominates the $\sim $ TeV observed
luminosity.

\begin{figure}
\epsscale{0.8}
\plotone{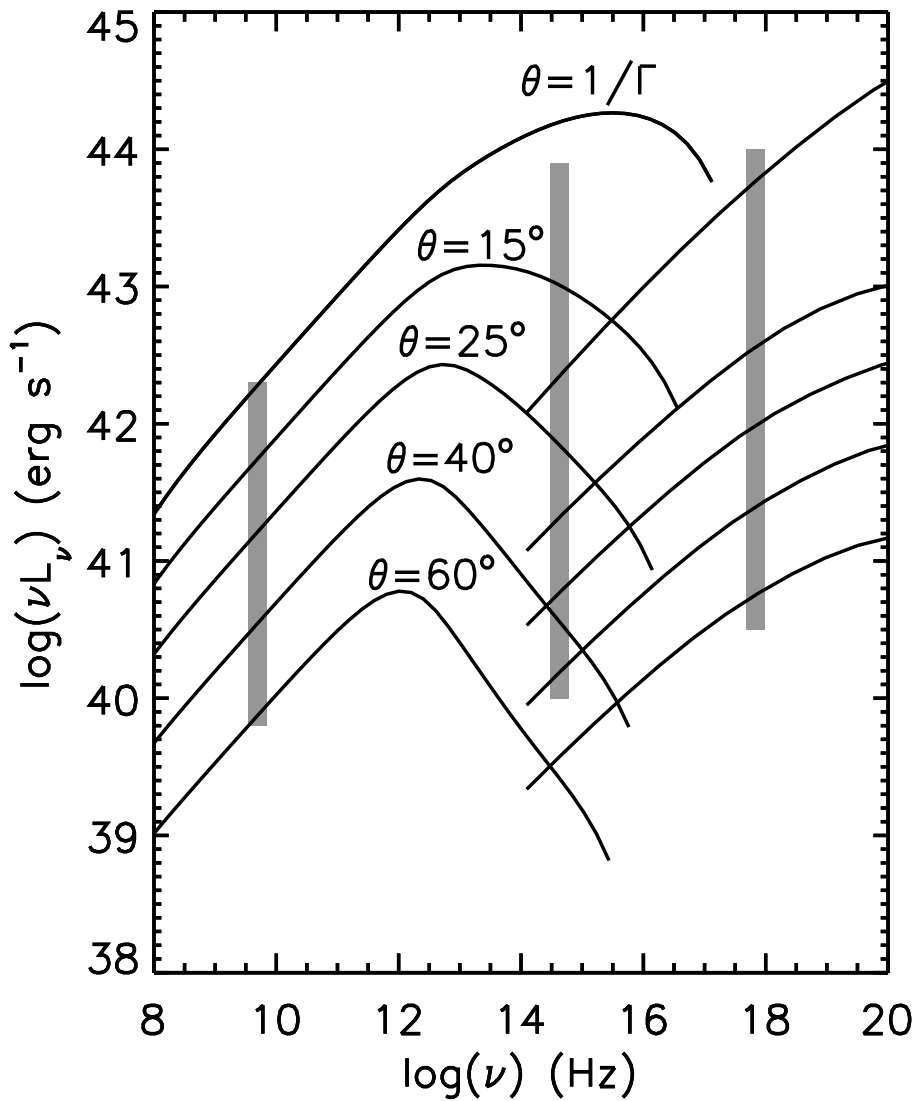}
\caption{The SED of a decelerating flow for a range of observing angles.
 The physical parameters are similar to the one shown in fig. \ref{fig1}, 
except from   the maximum electron energy which has been reduced to 
$\gamma_{max}=2\times 10^{5}$.
The shaded bars corresponds to the average luminosity difference
in radio, optical and X-rays, between the samples of Bls and FR I radio 
galaxies studied  by \cite{trussoni03}.}
\label{fig2}
\end{figure}

We now turn  to the problem of the unification of BLs with FR I
sources. \cite{chiaberge00} and \cite{trussoni03} compared  a sample of
FR I nuclei to BLs of similar extended radio power, which 
is believed to be non-beamed, and therefore orientation independent. 
They found that de-beaming 
the BL emission under the uniform velocity assumption by changing the 
observer's angle from $\theta = 1/\Gamma \simeq 4^{\circ}$ to $60^{\circ}$ 
leads to fluxes far smaller than those of FR Is.  In particular the average
BL to FR I nucleus luminosity ratio at radio, optical and X-ray bands
was found to be: $\log (L_{BL}/L_{FR\;I})_R \approx 2.4$,  $\log (L_{BL}/
L_{FR\;I})_{opt} \approx 3.9$,  $\log (L_{BL}/L_{FR\;I})_{X} \approx 3.5$.
In fig. \ref{fig2} we plot as vertical bars the luminosity separation of
BLs and FR Is according to  \cite{chiaberge00} and \cite{trussoni03}.
We also plot the SED of a decelerating flow with physical parameters similar 
to that in fig. 1, but with smaller value for $\gamma_{max}$ to
produce SED synchrotron peaks similar to those of the intermediate BLs 
that correspond in extended radio power to the FR Is \citep{chiaberge00,
trussoni03}. As can be seen, the luminosity change of the model 
SED at $\theta=60^{\circ}$ (FR I) and  $\theta=1/\Gamma$ (BL) reproduce 
relatively well the observed luminosity range.

\section{Discussion}

The need for additional seed photons in modeling the $\sim$ few TeV 
peak emission of the TeV blazars drives homogeneous SSC models to 
$\delta\gtrsim 50$, values in conflict with the presumed unification
between FR I's and BLs. 
The problem of the missing seed photons can be resolved if one considers 
a relativistic flow decelerating from $\Gamma_1\sim 15$ down to $\Gamma_2 
\sim $ a few: in this case UC  emission produces spectra that can easily  
provide the observed de-absorbed $E_p$ without the need to invoke 
values of $\delta$ greater than $\delta \simeq 15$.
Such decelerating flows 
are consistent with the low, possibly subluminal, speeds observed in the 
sub-pc scale jets of MKN 421 and MKN 501 without unreasonable alingment
requirements (for $\Gamma_2 =4$ and $\beta_{app} =1$ the corresponding value 
of the observing angle is $\theta = 2^{\circ}$). It also resolves the 
problem of FR I -- BL unification, which fails for flows with 
constant Lorentz factors even as low as $\Gamma \sim 10$.

The spatial separation of different frequencies seen in Fig. 1
has interesting consequences for the expected variability. 
 In homogeneous SSC   models
 a variation of the number
 of the injected  electrons produces a linear response in the
 synchrotron flux and a quadratic
one in the SSC flux. This is because  both the number 
of the electrons and the synchrotron energy density increase linerarly 
and  the SSC flux  increases
quadratically because  is proportional to their product.
For decelerating flows, fast variations 
(faster than the light crossing time of the separation between the  
X-ray emitting  region and the downstream 
region responsible for most of the  synchrotron seed
photons used to produce the TeV emission)  should result to 
 approximately linear variations of the TeV  relative to the X-ray flux.
This is because the freshly injected high energy electrons
UC scatter  mostly  synchrotron  photons 
produced  downstream before the injection, and  therefore
contribute an undisturbed photon  energy density. 

A physically plausible scenario for the flows we consider may be that 
suggested by Marscher (1999), according to which the energy dissipated
at the shock is converted into a non-thermal electron component,
whose radiative losses lead to the deceleration 
of the relativistic flow. In this case the deceleration length scale 
would be approximately equal to that of radiative losses, an assumption we 
have employed in our calculations. 
A similar scenario has been proposed for the hot spots 
of large scale jets, and it seems possible  that  relativistic and 
decelerating  flows  exist in different astrophysical environments
which exhibit similar characteristics, and in particular a stronger 
that anticipated high energy emission due to UC scattering.

\clearpage

\end{document}